\documentclass[preprint,showpacs,preprintnumbers,amsmath,amssymb]{revtex4}

\usepackage{graphicx}
\usepackage{dcolumn}
\usepackage{bm}

\begin{document}

\title{A Realistic Three-Dimensional Calculation of $^{3}$H Binding Energy}

\author{S. Bayegan}
\author{M.~R. Hadizadeh}
\email{hadizade@khayam.ut.ac.ir}
\author{M. Harzchi}

\affiliation{%
Department of Physics, University of Tehran, P.O.Box 14395-547,
Tehran, Iran
}%

\date{\today}

\begin{abstract}
A recently developed three-dimensional Faddeev integral equations
for three-nucleon bound state with two-nucleon interactions have
been solved in momentum space for Bonn-B potential.
\end{abstract}

\maketitle

\section{Introduction}

The study of the three-nucleon (3N) bound state with Faddeev
method based on an angular momentum decomposition, which includes
spin-isospin degrees of freedom, after truncation leads to a set
of a finite number of coupled equations in two variables for the
amplitudes and one needs a large number of partial waves to get
converged results \cite{Sammarruca-PRC46},\cite{Nogga-PLB409}. In
view of this very large number of interfering terms it appears
natural to give up such an expansion and work directly with vector
variables. On this basis recently three- and four-body bound
states have been studied in a Three-Dimensional (3D) approach
where as a simplification the spin-isospin degrees of freedoms
have been neglected in the first attempt
\cite{Liu-FBS33}-\cite{Hadizadeh-FBS40}. Considering the
spin-isospin is a major additional task, which will increase more
degrees of freedom into the states and will lead to a strictly
finite number of coupled equations. In this paper we implement
this task by including the spin-isospin degrees of freedom in 3N
bound state formalism. We formulate the Faddeev equations with NN
interactions as function of vector Jacobi momenta, specifically
the magnitudes of the momenta and the angles between them. We
obtain a strictly finite number of coupled three-dimensional
integral equations in 3 variables for the amplitudes which greatly
simplifies the calculations without using a PW decomposition. We
solve the coupled Faddeev integral equations for calculation of
Triton binding energy with Bonn-B potential. The input to our
calculations is the two-body $t$-matrix which has been calculated
in an approach based on a Helicity representation and depends on
the magnitudes of the initial and final momenta and the angle
between them \cite{Fachruddin-PRC62}.

\section{Faddeev Equations in a Realistic 3D Approach}

The bound state of three pairwise-interacting nucleons is
described by Faddeev equation:

\begin{eqnarray}
|\psi\rangle =G_{0}t_{12}P |\psi\rangle \label{eq.1}
\end{eqnarray}

Here the free 3N propagator is given by $G_{0}=(E-H_{0})^{-1}$,
and $H_{0}$ stands for the free Hamiltonian. $t_{12}$ and $P$ are
the two-body transition and permutation operators. The total
anti-symmetrized wave function $|\Psi\rangle$ of 3N system is
composed of three Faddeev components as $|\Psi\rangle=(1+P)
|\psi\rangle$. The antisymmetry property of $|\psi\rangle$ under
exchange of interacting particles $1$ and $2$ guarantees that
$|\Psi\rangle$ is totally antisymmetric. In order to solve
equation (\ref{eq.1}) in momentum space we introduce the 3N basis
states in a 3D formalism as, \cite{Bayegan-PRC}:
\begin{eqnarray}
  |\, {\bf p} \, {\bf q} \,\,  \alpha  \, \rangle \equiv | \,
 {\bf p} \, {\bf q} \,\, (s_{12} \,\,
 \frac{1}{2}) S \, M_{S} \,  \,\, (t_{12} \,\,
 \frac{1}{2}) T \, M_{T} \,  \rangle
 \label{eq.2}
\end{eqnarray}

As indicated in Eq.~(\ref{eq.2}) the basis states involve two
standard Jacobi momenta and consequently the angular dependence
explicitly appears in these Jacobi vector variables, whereas in a
standard PW approach the angular dependence leads to two orbital
angular momentum quantum numbers, i.e. $l_{12}$ and $l_{3}$.
Therefore we couple the spin quantum numbers $s_{12}$ and $s_{3}$
to the total spin $S$ and its third component $M_{S}$ as: $| \,
(s_{12} \,\, s_{3})S \, M_{S} \, \rangle$. For the isospin quantum
numbers similar coupling scheme leads to total isospin $T, \,
M_{T}$. For evaluating the Faddeev equation, Eq. (\ref{eq.1}), we
need to evaluate the matrix elements of two-body $t$-matrix and
permutation operators. They have been evaluated in detail in Ref.
\cite{Bayegan-PRC} as:

\begin{eqnarray}
\langle \, {\bf p}\,{\bf q}\, \alpha \, |\psi\rangle &=&
\frac{1}{{E-\frac{p^{2}}{m} -\frac{3q^{2}}{4m}}} \, \sum_{\gamma'}
\, \int d^{3}q' \, \sum_{\gamma''} \, g_{\alpha \gamma''} \,
 \delta_{m''_{s_{3}} m'_{s_{1}}} \, \delta_{m''_{t_{3}}
m'_{t_{1}}} \nonumber
\\*  &\times& \,\, _{a}\langle{\bf p}\, m''_{s_{1}}
m''_{s_{2}} \, m''_{t_{1}} m''_{t_{2}} |t(\epsilon)
|\frac{-1}{2}{\bf q}-{\bf q}' \, m'_{s_{2}} m'_{s_{3}} \,
m'_{t_{2}} m'_{t_{3}} \rangle_{a} \, \nonumber
\\* &\times&  \sum_{\alpha'} \, g_{\gamma' \alpha'} \langle{\bf q}+\frac{1}{2}{\bf q}'
\,\, {\bf q}' \, \alpha'|\psi\rangle
 \label{eq.3}
 \end{eqnarray}

 where $\gamma'(\gamma'')$ are the spin-isospin part of free 3N basis
 states and $g_{\alpha \gamma''}$ are Clebsch-Gordan coefficients.
For derivation of Eq. (\ref{eq.3}) the anti-symmetry property of
the Faddeev component as well as the physical representation of
the two-body t-matrix \cite{Fachruddin-PRC62} are used. In order
to solve this equation directly without employing PW projections,
we choose the spin polarization direction parallel to the $z$-axis
and express the momentum vectors in this coordinate system.

Since the angular momentum quantum numbers don't appear explicitly
in our formalism, therefore the number of coupled equations which
are fixed according to the spin-isospin states are reduced. This
is an indication that the present formalism automatically consider
all partial waves without any truncation on the space part.
Considering the spin-isospin degrees of freedom for both $^{3}H$
and $^{3}He$ states yields the same number of coupled equations
and it leads to 8, 12, 16 and 24 coupled equations for different
combinations of total spin-isospin states $S-T$:
$(\frac{1}{2}-\frac{1}{2})$,
$(\frac{1}{2}-\frac{1}{2},\frac{3}{2})$,
$(\frac{1}{2},\frac{3}{2}-\frac{1}{2})$ and
$(\frac{1}{2},\frac{3}{2}-\frac{1}{2},\frac{3}{2})$ respectively.
In a standard PW approach the infinite set of coupled integral
equations is truncated in actual calculations at sufficiently high
values of angular momentum quantum numbers. If one assumes that
the NN $t$-matrix acts up to $j_{12}^{max}=1, 2, 3, 4$ and $5$
then the number of channels will be 5, 18, 26, 34 and 42, while
the total isospin is restricted to $T=\frac{1}{2}$
\cite{Machleidt-ANP19}.

\section{$^{3}$H Binding Energy}

In order to test our realistic 3D formalism for 3N bound state we
solve the eight coupled Faddeev three-dimensional integral
equations corresponding to total spin-isospin states
$(\frac{1}{2}-\frac{1}{2})$ for $^{3}$H case . We calculate Triton
binding energy with Bonn one-boson-exchange (OBE) potential in the
parametrization of Bonn-B \cite{Machleidt-ANP19} and in an
operator form which can be incorporated in 3D formalism
\cite{Fachruddin-PRC62}.

As shown in table \ref{table1} our calculations for Bonn-B NN
potential in 3D approach yield the value $-8.15\, [MeV]$ for
$^{3}$H binding energy, which is in good agreement with the
converged value $-8.14 \, [MeV]$ of Faddeev calculations in PW
scheme for $j_{12}^{max}=4$. As we can see from this comparison
our result provide the same accuracy while the numerical procedure
is actually easier to implement.

\begin{table}[hbt]
\caption {Triton binding energy in 3D approach in comparison with
the PW result.}
\begin{tabular}{cccccccc}
\hline \hline
Method  &&&&&  $E_{t}$ [MeV]  \\
\hline
 PW ($j_{12}^{max}=4$) \cite{Sammarruca-PRC46}, \cite{Machleidt-ANP19}, \cite{Witala-PRC43}  &&&&& -8.14 \\
\hline\hline
3D  &&&&&  -8.15  \\
\hline\hline
\end{tabular}
\label{table1}
\end{table}


\begin{thebibliography}{9}

\bibitem{Sammarruca-PRC46} F. Sammarruca, D. P. Xu and R. Machleidt, {\it Phys. Rev.} {\bf C46}, 1636 (1992).

\bibitem{Nogga-PLB409} A. Nogga, D. H\"{u}ber, H. Kamada and W. Gl\"{o}ckle, {\it Phys. lett.} {\bf B409}, 19 (1997).

\bibitem{Machleidt-ANP19} R. Machleidt, {\it Adv. Nucl. Phys.} {\bf 19}, 189 (1989).

\bibitem{Witala-PRC43} H. Witala, W. Gl\"{o}ckle and H. Kamada, {\it Phys. Rev.} {\bf C43}, 1619 (1991).

\bibitem{Liu-FBS33} H. Liu, Ch. Elster, W. Gl\"{o}ckle, {\it Few-Body Systems} {\bf 33},
241 (2003).

\bibitem{Hadizadeh-FBS40} M. R. Hadizadeh and S. Bayegan, {\it Few-Body Systems} {\bf 40},
171 (2007).

\bibitem{Fachruddin-PRC62} I. Fachruddin, Ch. Elster, and W. Gl\"{o}ckle, {\it Phys. Rev.} {\bf C62}, 044002 (2000).

\bibitem{Bayegan-PRC} S. Bayegan, M. R. Hadizadeh and M. Harzchi, {\it submitted to Phys. Rev.} {\bf C}.


\end{thebibliography}
\end{document}